\documentstyle[11pt,epsfig]{article}

\begin{document}

\setcounter{page}{1}

\title{$SU(2)$ Dirac-Yang-Mills quantum mechanics of spatially constant quark and gluon fields}
\author{H.-P. Pavel \\
        Institut f\"ur Kernphysik, TU Darmstadt,
        Schlossgartenstr. 9,
        D-64289 Darmstadt, Germany\footnote{email: hans-peter.pavel@physik.tu-darmstadt.de}\\
        and \\
        Bogoliubov Laboratory of Theoretical Physics,
        Joint Institute for Nuclear Research, Dubna, Russia\\
        }
\date{Mai 3, 2011}
\maketitle
\begin{abstract}
The quantum mechanics of spatially constant $SU(2)$ Yang-Mills- and Dirac-fields minimally coupled to
each other is investigated as the strong coupling limit of 2-color-QCD.
Using a canonical transformation of the quark and gluon fields,
which Abelianises the Gauss law constraints to be implemented, the corresponding unconstrained
Hamiltonian and total angular momentum are derived. In the same way as
this reduces the colored spin-1 gluons to unconstrained colorless
spin-0 and spin-2 gluons, it reduces the colored spin-$\frac{1}{2}$ quarks to
unconstrained colorless spin-0 and spin-1 quarks. These however continue
to satisfy anti-commutation relations and hence the Pauli-exclusion principle.
The obtained unconstrained Hamiltonian is then rewritten into a
form, which separates the rotational from the scalar degrees of freedom.
In this form the low-energy spectrum can be obtained with high accuracy.
As an illustrative example, the spin-0 energy-spectrum of the quark-gluon system
is calculated for massless quarks of one flavor. It is found, that only for the case of 4 reduced
quarks (half-filling) satisfying the boundary condition of particle-antiparticle
C-symmetry, states with energy lower than for the pure-gluon case are obtained.
These are the ground state, with an energy about $20\%$ lower than for the pure-gluon case
and the formation of a quark condensate,
and the sigma-antisigma excitation with an energy about a fifth of that
of the first glueball excitation.
\end{abstract}


\section{\large\bf Introduction}

For a complete description of the physical properties of low-energy QCD,
such as color confinement, chiral symmetry breaking, the formation of condensates and flux-tubes,
and the spectra and strong interactions of hadrons,
it might be advantageous if one could first reformulate QCD in terms of gauge invariant
dynamical variables, before applying any approximation schemes (see e.g.\cite{Christ and Lee}).
Using a canonical transformation of the dynamical variables,
which Abelianises the Non-Abelian Gauss-law constraints,
such a reformulation has been achieved for pure $SU(2)$ Yang-Mills theory
on the classical \cite{GKMP,KP1,KMPR} and on the quantum level \cite{pavel2}.
The resulting unconstrained
$SU(2)$ Yang-Mills Hamiltonian admits a systematic strong-coupling expansion in powers of $\lambda=g^{-2/3}$,
equivalent to an expansion in the number of spatial derivatives.
The leading order term in this expansion constitutes
the unconstrained Hamiltonian of $SU(2)$ Yang-Mills quantum mechanics of spatially constant gluon fields
\cite{Luescher and Muenster}-\cite{pavel1}, for which the low-energy spectra can be calculated with high accuracy.
Subject of the present work is its generalisation to the case
of $SU(2)$ Dirac-Yang-Mills quantum mechanics of quark and gluon fields\footnote{
This is different from the works \cite{van Baal,Michael} where the effect of dynamical quarks, satisfying
anti-periodic boundary conditions, on the spatially constant gluon fields in a small volume has been investigated.
Here we study spatially constant quark and gluon fields in large volumes.
}.
First steps in this direction on the classical level have been done in \cite{GKMP}.

Under the supposition of the spatial homogeneity of the fields,
the Lagrangian of 2-color-QCD reduces to\footnote{
Everywhere in the paper we put the spatial volume $V= 1$.
 As result the coupling constant $g$ becomes dimensionful
 with $g^{2/3}$ having the dimension of energy. The volume dependence
 can be restored in the final results by replacing $g^2$ with $g^2/V$. }
\begin{equation}
\label{hl}
L={1\over 2}\left(\dot{A}_{ai}-g\epsilon_{abc}A_{b0} A_{ci}\right)^2 -{1\over 2} B_{ai}^2(A)
   +{i\over 2}\left(\psi^*_{\alpha r}\,\dot{\psi}_{\alpha r}
                    -\dot{\psi}_{\alpha r}^* \,\psi_{\alpha r}\right)
   - g A_{a0}\, \rho_a(\psi)
   + g A_{ai}\, j_{i a}(\psi)
   -m\, \overline{\psi} \psi~,
\end{equation}
with  the magnetic field
$B_{ai}(A)= (1/2)g\epsilon_{abc}\epsilon_{ijk}A_{bj}A_{ck}$, the Lorentz scalar
$\overline{\psi} \psi=\psi^*_{\alpha r} \beta_{rs} \psi_{\alpha s}$, and the color-densities and -currents
(using the Pauli matrices $\tau_a\, (a=1,2,3)$)
\begin{equation}
\rho_a(\psi):={1\over 2}\ \psi_{\alpha r}^*
          \,\left({\mathbf{\tau}}_a\right)_{\alpha\beta}\psi_{\beta r}~,
\quad\quad
j_{i a}(\psi):={1\over 2}\ \!\psi_{\alpha r}^*
\left({\mathbf{\tau}}_a\right)_{\alpha\beta}\left(\alpha_i\right)_{rs}\psi_{\beta s}~.
\end{equation}
For simplicity we limit ourselves here to one quark flavor.
The local $SU(2)$ gauge invariance of the original 2-color-QCD action
reduces to the symmetry under the $SU(2)$ transformations $U(\omega(t))$ of the fermion fields,
local in time,
\begin{equation}
\psi^{\omega}_{\alpha r}(t)=U(\omega(t))_{\alpha\beta}\ \psi_{\beta r}(t)~,
\end{equation}
and the corresponding $SO(3)$ transformations
$
O(\omega(t))_{ab} =
{1\over 2}{\rm Tr}\left(U^{-1}(\omega(t)){\mathbf{\tau}}_a U(\omega(t)){\mathbf{\tau}}_a\right)
$
of the gauge fields,
\begin{eqnarray}
 A^{\omega}_{a0}(t)\!\!\!&=&\!\!\!
O(\omega(t))_{ab}A_{b0}(t) -\frac{1}{2g}\
\epsilon_{abc}\left(O(\omega(t))\dot O(\omega(t)) \right)_{bc}\,,
\nonumber\\
 A^{\omega}_{ai}(t)\!\!\!&=&\!\!\!
O(\omega(t))_{ab}A_{bi}(t)~.
\label{tr}
\end{eqnarray}
The rotational invariance of the original Yang-Mills action reduces to the global spatial rotations
\begin{equation}
A^{\chi}_{ai}=A_{aj}R(\chi)_{ji}~,\quad\quad
\psi^{\chi}_{\alpha r}=\Lambda(\chi)_{rs}\psi_{\alpha s}~,
\end{equation}
where the $4\times 4$ Dirac rotation matrix $\Lambda(\chi)$ is related to $R(\chi)$ via
$
R(\chi)_{ij} =
{1\over 2}{\rm Tr}\left(\Lambda^{-1}(\chi)\gamma_i \Lambda(\chi)\gamma_j \right).
$
The canonical Hamiltonian obtained from (\ref{hl}) via Legendre-transformation reads
\begin{equation}
H_C={1\over 2}\Pi_{ai}\Pi_{ai}+{1\over 2} B_{ai}^2(A)
   + g A_{a0} \left(\epsilon_{abc} A_{bi}\Pi_{ci} + \rho_a(\psi)\right)
   - g A_{ai}\, j_{i a}(\psi)
   +m\, \overline{\psi} \psi~,
\end{equation}
where $\Pi_{ai}$ are the momenta canonical conjugate to the spatial components $A_{ai}$.

In the constrained Hamiltonian formulation (see e.g.\cite{Christ and Lee})
the time dependence of the gauge transformations (\ref{tr}) is exploited
to put the Weyl gauge
\begin{equation}
A_{a0} = 0~,\quad\quad a=1,2,3,
\end{equation}
on the remaining dynamical degrees of freedom $A_{ai}$, $\Pi_{ai}$, $\psi_{\alpha r}$ and $\psi^*_{\alpha r}$
 are quantized in the Schr\"odinger functional approach
by imposing the equal time commutation  relations
\begin{equation}
\label{comm}
\Pi_{ai} \rightarrow -i\partial/\partial A_{ai}~:\quad\quad
[\Pi_{ai}\,, A_{bj}]=-i\delta_{a b}\delta_{ij}~,
\end{equation}
and anti-commutation relations
\begin{equation}
\label{anticomm}
\psi^*_{\alpha r}  \rightarrow   \psi^\dag_{\alpha r}~:\quad\quad
\{ \psi_{\alpha r},\psi^\dag_{\beta s}\}=\delta_{\alpha\beta}\delta_{rs}~,
\quad\quad
\{ \psi_{\alpha r},\psi_{\beta s}\}=0~.
\end{equation}
where the quark and gluon field operators commute
\begin{equation}
\label{commApsi}
[A_{ai}\,, \psi_{\alpha r}]=0~,\quad\quad [\Pi_{ai}\,, \psi_{\alpha r}]=0~.
\end{equation}
The physical states $\Phi$ have to satisfy both the Schr\"odinger equation
and the three Gauss law constraints
\begin{eqnarray}
H\Phi &=&
 \left[{1\over 2}\left(\frac{\partial}{\partial A_{ai}}\right)^2+{1\over 2}B_{ai}^2(A)
- A_{ai}\, j_{i a}(\psi) +m\, \overline{\psi} \psi\right]\Phi=E\Phi~,
\quad\quad \\
G_a\Phi &=& \left[-i\epsilon_{abc}A_{bi}\frac{\partial}{\partial A_{ci}}
+\rho_a(\psi)\right]\Phi=0~,
\quad a=1,2,3~.\label{G_a}
\end{eqnarray}
The $G_a$ are the generators of the residual time independent gauge transformations,
satisfying $[G_a,H]=0$ and $[G_a,G_b]=i\epsilon_{abc}G_c$.
Furthermore $H$ commutes with the angular momentum operators
\begin{eqnarray}
\label{constrainedJ}
J_i  =  -i\epsilon_{ijk}A_{aj} {\partial\over\partial A_{ak}} + \Sigma_i(\psi) ~,
\quad i=1,2,3~,
\end{eqnarray}
with the quark-spin
\begin{equation}
\Sigma_i(\psi) :={i\over 8}\,\epsilon_{ijk}\, \psi^\dag_{\alpha r}
              [\gamma_j,\gamma_k]_{rs}\,\psi_{\alpha s}~.
\end{equation}
The matrix element of an operator $O$ is given in the Cartesian form
\begin{equation}
\langle \Phi'| O|\Phi\rangle\
\propto
\int dA\ d\overline{\psi} d\psi\
\Phi'^*(A,\overline{\psi},\psi)\, O\, \Phi(A,\overline{\psi},\psi)~.
\end{equation}
For carrying out quantum mechanical calculations it is desirable to
have a corresponding unconstrained Schr\"odinger equation and to find its eigenstates
in an effective way with high accuracy at least for the lowest states.

\section{\large\bf Unconstrained Dirac-Yang-Mills Hamiltonian}

The local symmetry transformation (\ref{tr}) of the gauge potentials
\( A_{ai} \) prompts us with the set of coordinates in terms of which the
separation of  the  gauge degrees of freedom occurs.
This can be achieved \cite{GKMP} using the polar decomposition for arbitrary
\(3\times 3\) quadratic matrices, and the new fermionic variables
$\psi^\prime_s$
\begin{eqnarray}
\label{eq:pcantr}
A_{ai} \left(q, S \right)
&=& O_{ak}\left( q \right) S_{ki}~,
\\
\label{eq:pcantrferm}
\psi_\alpha\left(q, \psi^\prime \right)&=& U_{\alpha\beta}\left( q \right) \psi^\prime_{\beta}
\end{eqnarray}
with the orthogonal matrix \( O (q)  \), parametrized by the three angles \(q_i\),
which is the adjoint representation of the unitary $2\times 2$ matrix $U(q)$
\begin{equation}
O_{ab}(q)={1\over 2}\mbox{Tr}\left(U^{-1}(q)\tau_a U(q)\tau_b\right)~.
\end{equation}
and the positive definite, symmetric \(3\times 3\)  matrix \( S \). The decomposition
(\ref{eq:pcantr}) is unique and corresponds to the symmetric gauge \cite{KP1,KMPR}
\begin{equation}
\chi_i(A)=\epsilon_{ijk}A_{jk}=0~.
\end{equation}
Preserving the canonical commutators (\ref{comm}) and (\ref{commApsi})
one obtains the expressions for the old canonical momenta in terms of the new variables
\begin{equation}
\label{canmom}
-i{\partial\over\partial A_{ai}}=O_{ak}\left(q\right)\left[-i{\partial\over\partial S_{ki}}+\epsilon_{kil}
\gamma^{-1}_{ls}(S)\left(-i\Omega^{-1}_{sj}(q){\partial\over\partial q_{j}}
+\rho_s(\psi^{\prime})
-i\epsilon_{smn}S_{bm} {\partial\over\partial S_{bn}}\right)\right]
\end{equation}
where $\Omega_{jm}(q) \equiv (1/2) \epsilon_{mkl}
\left[O^T (q)\partial O\left(q\right)/\partial q_j\right]_{kl}$.
The Jacobian  is
$|\partial(A_{ai})/\partial(q,S)|\
\propto \det\Omega(q)\prod_{i<j}\left(\phi_i+\phi_j\right)$,
where $\phi_1,\phi_2,\phi_3$ are the eigenvalues of $S$.
The variables \( S \) and \(\partial/\partial S  \) make no contribution to the
Gauss law operators
\begin{equation}
G_a = -iO_{as}(q) \Omega^{-1}_{\ sj}(q){\partial\over\partial q_j}~.
\end{equation}
Hence, assuming the invertibility of  the matrix
$\Omega$, the non-Abelian Gauss laws (\ref{G_a}) can be replaced
by an equivalent set of Abelian constraints
\begin{equation}
G_a\Phi =0 ~,\quad a=1,2,3~ \quad\quad \Leftrightarrow \quad\quad {\partial \Phi\over\partial q_i}  = 0~,\quad i=1,2,3.
\quad ({\rm Abelianisation})
\end{equation}
and the unconstrained Hamiltonian and total angular momentum of
$SU(2)$ Dirac-Yang-Mills quantum mechanics read
\begin{eqnarray}
\label{unconstrainedH}
 H  &=& {1\over 2}
\sum_{m,n} \Bigg[
-\left(\frac{\partial}{\partial S_{mn}}\right)^2
-\left[\gamma^{-1}_{mn}(S)-\delta_{mn}\mbox{tr}(\gamma^{-1}(S))\right]
    \frac{\partial}{\partial S}_{mn}
+B_{mn}^2(S)
\nonumber\\  && \quad\quad\quad\quad
-g\psi^{\prime \dag}\alpha_m S_{mn}\tau_n\psi^{\prime}
+{1\over 2}\left(J_m-i K^Q_m\right)\gamma^{-2}_{mn}(S)
\left(J_n-i K^Q_n\right)
\Bigg] +{1\over 2}m\, \overline{\psi^{\prime}}\psi^{\prime}~.
\\
\label{unconstrainedJ}
J_i &=& -2i\epsilon_{ijk}S_{aj}\ {\partial\over\partial S_{ak}} + J^Q_i~,\quad\quad [J_i,H]=0~,
\end{eqnarray}
in terms of the reduced variables, where
$\gamma_{ik}(S) : = S_{ik} -  \delta_{ik}\mbox{tr} S$,
and  $J^Q_i$ and $K^Q_i$ are the quark operators
\begin{eqnarray}
J^Q_i :=
\Sigma_i(\psi^{\prime})+\rho_i(\psi^{\prime})
~,\quad\quad
K^Q_i :=
-i\Big(\Sigma_i(\psi^{\prime})-\rho_i(\psi^{\prime})\Big)
  ~,\quad\quad i=1,2,3,
\end{eqnarray}
satisfying
\begin{equation}
\label{JKalgebra}
[J^Q_i,J^Q_j]=i\epsilon_{ijk}J^Q_k~,\quad\quad
[J^Q_i,K^Q_j]=i\epsilon_{ijk}K^Q_k~,\quad\quad
[K^Q_i,K^Q_j]=-i\epsilon_{ijk}J^Q_k~.
\end{equation}
The matrix element of a physical operator O is given by
\begin{equation}
\langle \Phi'| O|\Phi\rangle\
\propto
\int dS\ d\overline{\psi^\prime} d\psi^\prime\
\Big[\prod_{i<j}\left(\phi_i+\phi_j\right)\Big]
\Phi'^*(S,\overline{\psi^\prime},\psi^\prime)\, O\, \Phi(S,\overline{\psi^\prime},\psi^\prime)~.
\end{equation}
The classical analog of the unconstrained Hamiltonian (\ref{unconstrainedH}) in some equivalent
form has already been obtained in \cite{GKMP}, the expression (\ref{unconstrainedJ})
of the unconstrained total angular momentum, however, is new.
Note that the extra factor of $2$ in the gluonic part and the additional term $\rho_i(\psi^{\prime})$ in the
quark part $J^Q_i$ of the physical total spin (\ref{unconstrainedJ}),
in comparison with the constrained form (\ref{constrainedJ}), originate from the anti-symmetric
part of the momenta (\ref{canmom}) and have important consequences.

The variables $S$ transform under spatial rotations as symmetric tensor field,
\begin{equation}
\label{trafoJgluon}
[J_i,S_{mn}]=i\left(\epsilon_{imj}S_{jn}+\epsilon_{inj}S_{mj}\right)~,\quad\quad i=1,2,3,
\end{equation}
which can be decomposed into spin-0 and spin-2 components (using Clebsch-Gordan coefficients)
\begin{equation}
S_{ik}=C_{1i\ 1k}^{2 A}\ S^{\!(2)}_{A}
         +\frac{1}{\sqrt{3}}\delta_{ik}\ S^{\!(0)}
\end{equation}

In the Weyl-representation, the quark-spin operators $J^Q_i$ can be written
\begin{equation}
\label{S4}
J^Q_i =:\psi^{\prime \dag}_L \widehat{S}_i^{(4)}\psi^{\prime}_L+
   \psi^{\prime \dag}_R \widehat{S}_i^{(4)}\psi^{\prime}_R~,
\quad\quad
\widehat{S}_i^{(4)}={1 \over 2}\left(\tau_i \otimes 1_2 + 1_2 \otimes \tau_i\right)~,
\quad\quad
[\widehat{S}_i^{(4)},\widehat{S}_j^{(4)}]=i\epsilon_{ijk}\widehat{S}_k^{(4)}~,
\end{equation}
with the three $4\times 4$ matrices $\widehat{S}_i^{(4)}$
generate $SO(3)$ spatial rotations
and read explicitly
\begin{equation}
\label{S4a}
\widehat{S}_1^{(4)}={1 \over 2}
\left(\begin{array}{cccc}
0&  1   &   1    &0   \cr
1&  0   &   0    &1   \cr
1&  0   &   0    &1   \cr
0&  1   &   1    &0
\end{array}\right),
\quad
\widehat{S}_2^{(4)}={1 \over 2}
\left(\begin{array}{cccc}
0&  -i  &   -i   &0   \cr
i&  0   &   0    &-i  \cr
i&  0   &   0    &-i  \cr
0&  i   &   i    &0
\end{array}\right),
\quad
\widehat{S}_3^{(4)}=
\left(\begin{array}{cccc}
1&  0  &   0   &0   \cr
0&  0   &   0    &0  \cr
0&  0   &   0    &0  \cr
0&  0   &   0    &-1
\end{array}\right)~.
\end{equation}
It follows that the combinations
\begin{eqnarray}
\psi_{L}^{\prime (0)}\!\! :=\!\!\frac{1}{\sqrt{2}}\left(-\psi_{L2}^{\prime}+\psi_{L3}^{\prime}\right),
\ \
\left(\psi_{L1}^{\prime (1)},\psi_{L2}^{\prime (1)},\psi_{L3}^{\prime (1)} \right)\!\! := \!\!
\Big(\! -\frac{1}{\sqrt{2}}\left(\psi_{L1}^{\prime}-\psi_{L4}^{\prime}\right),
 -\frac{i}{\sqrt{2}}\left(\psi_{L1}^{\prime}+\psi_{L4}^{\prime}\right),
\frac{1}{\sqrt{2}}\left(\psi_{L2}^{\prime}+\psi_{L 3}^{\prime}\right)\!\!\Big),
\end{eqnarray}
of the reduced quark fields $\psi^{\prime}_L$,
and analogously for the right-handed components,
transform as spin-0 and spin-1 fields
\begin{equation}
\label{trafoJferm}
[J_i,\psi_{L(R)}^{\prime (0)}]\equiv[J^Q_i,\psi_{L(R)}^{\prime (0)}]=0~,\quad\quad
[J_i,\psi_{L(R)\ j}^{\prime (1)}]\equiv[J^Q_i,\psi_{L(R)\ j}^{\prime (1)}]
=i\epsilon_{ijk}\psi_{L(R)\ k}^{\prime (1)}
~,\quad\quad i=1,2,3.
\end{equation}
Hence implementation of the Gauss law constraints reduces the original spin-1/2 quark fields $\psi$
to the unconstrained quark fields $\psi^\prime$ carrying integer spin-0 and spin-1,
just as the original spin-1 gluon fields $A$ are reduced to
the corresponding unconstrained spin-0 and spin-2 gluon fields $S$.
It is however to notice that due to the unitarity of transformation (\ref{eq:pcantrferm}),
the unconstrained quarks
continue to satisfy anti-commutation relations and hence the Pauli exclusion principle\footnote{
Note, that there is no spin-statistics theorem for non-local theories.
}.

The anti-Hermitean quark operators $K^Q_i$ can be written as
\begin{equation}
\label{T4}
K^Q_i =:\psi^{\prime \dag}_L \widehat{T}_i^{(4)}\psi^{\prime}_L+
   \psi^{\prime \dag}_R \widehat{T}_i^{(4)}\psi^{\prime}_R~,
\quad\quad
\widehat{T}_i^{(4)}=-{i \over 2}\left(\tau_i \otimes 1_2 - 1_2 \otimes \tau_i\right)~,
\quad\quad
[\widehat{T}_i^{(4)},\widehat{T}_j^{(4)}]=-i\epsilon_{ijk}\widehat{S}_k^{(4)}~,
\end{equation}
with the three $4\times 4$ matrices $\widehat{T}_i^{(4)}$
\begin{equation}
\label{T4a}
\widehat{T}_1^{(4)}={i \over 2}
\left(\begin{array}{cccc}
0&  1   &   -1    &0   \cr
1&  0   &   0    &-1   \cr
-1&  0   &   0    &1   \cr
0&  -1   &   1    &0
\end{array}\right),
\quad
\widehat{T}_2^{(4)}={1 \over 2}
\left(\begin{array}{cccc}
0&  1  &   -1   &0   \cr
-1&  0   &   0    &-1  \cr
1&  0   &   0    &1  \cr
0&  1   &   -1    &0
\end{array}\right),
\quad
\widehat{T}_3^{(4)}=-i
\left(\begin{array}{cccc}
0&  0  &   0   &0   \cr
0&  1   &   0    &0  \cr
0&  0   &   -1    &0  \cr
0&  0   &   0    &0
\end{array}\right)~,
\end{equation}
which generate boosts
\begin{equation}
[K^Q_i,\psi_{L(R)}^{\prime (0)}]=-i\psi_{L(R)\ i}^{\prime (1)}~,\quad\quad
[K^Q_i,\psi_{L(R)\ j}^{\prime (1)}]=-i\delta_{ij}\psi_{L(R)}^{\prime (0)}
~,\quad\quad i=1,2,3,
\end{equation}
on the left(right)-handed 4-vector
$(\psi_{L(R)}^{\prime (0)},
\ \psi_{L(R)\ 1}^{\prime (1)},\ \psi_{L(R)\ 2}^{\prime (1)},\ \psi_{L(R)\ 3}^{\prime (1)})$
of reduced quark fields.
For example, we easily verify the Lorentz invariant
\begin{eqnarray}
[K^Q_i,\overline{\psi^{\prime}} \psi^\prime]=
[K^Q_i,\left(\psi^{\prime (0)\dag}_L\psi^{\prime (0)}_R+\psi^{\prime (0)\dag}_R\psi^{\prime (0)}_L\right)
+\sum_{i=1}^3\left(\psi^{\prime (i)\dag}_L\psi^{\prime (i)}_R+\psi^{\prime (i)\dag}_R\psi^{\prime (i)}_L\right)]=0~.
\end{eqnarray}

\section{\large\bf Unconstrained Hamiltonian in terms of
rotational and scalar degrees of freedom}

\subsection{\bf Transformation to
rotational and scalar degrees of freedom}

A more transparent form for the unconstrained Dirac-Yang-Mills
Hamiltonian (\ref{unconstrainedH}), maximally separating the rotational from the rotation
invariant degrees of freedom,
can be obtained using the transformation properties (\ref{trafoJgluon}) and
(\ref{trafoJferm}) of the canonical fields $S$ and $\psi^\prime$ under spatial rotations
generated by the physical total spin (\ref{unconstrainedJ}).
We limit ourselves in this work to the case of principle orbit configurations
of non-coinciding eigenvalues $\phi_1,\phi_2,\phi_3>0$ of the positive definite symmetric matrix $S$,
which without loss of generality can be taken as
\begin{equation}
\label{range}
0<\phi_1<\phi_2<\phi_3<\infty~,
\end{equation}
(not considering singular orbits where two or more eigenvalues coincide)
and perform a principal-axes transformation
\begin{eqnarray}
\label{patransf}
S(\chi,\phi)  =  R(\chi)
\left(\begin{array}{ccc}
\phi_1&0&0\cr
0&\phi_2&0\cr
0&0&\phi_3
\end{array}\right)
 R^{T}(\chi)~,
\end{eqnarray}
with the \( SO(3)\) matrix  \({R}(\chi)\) parametrized
by the three Euler angles $\chi\equiv(\alpha,\beta,\gamma)$,
\begin{equation}
R(\chi)=\exp(-i\alpha \widehat{S}_z) \exp(-i\beta \widehat{S}_y)
\exp(-i\gamma \widehat{S}_z)~,
\end{equation}
with the \( SO(3)\) generators $(\widehat{S}_i)_{jk}=-i\epsilon_{ijk}$.
On the fermion fields in Weyl representation, the transformation
\begin{equation}
\label{patransfferm}
\psi^\prime (\chi,\widetilde{u}_L,\widetilde{v}_R)
={\psi_L^\prime(\chi,\widetilde{u}_L) \choose \psi_R^\prime(\chi,\widetilde{v}_R)}
={\widetilde{U}(\chi)\widetilde{u}_L \choose \widetilde{U}(\chi)\widetilde{v}_R}
\end{equation}
is performed, with the  $U(4)$ matrix (${\rm det}\,\widetilde{U}(\chi)=-i$)
\begin{equation}
\label{U}
\widetilde{U}(\chi)= \widetilde{R}(\chi)\ \widetilde{U}_0 =
\left(\begin{array}{cccc}
0           & D^{(1)}_{1\ 1-}(\chi)          &
D^{(1)}_{1\ 1+}(\chi)          & D^{(1)}_{1\ 0}(\chi) \cr
 -1/\sqrt{2} & D^{(1)}_{0\ 1-}(\chi)/\sqrt{2} &
D^{(1)}_{0\ 1+}(\chi)/\sqrt{2} & D^{(1)}_{0\ 0}(\chi)/\sqrt{2}\cr
 1/\sqrt{2} & D^{(1)}_{0\ 1-}(\chi)/\sqrt{2} &
D^{(1)}_{0\ 1+}(\chi)/\sqrt{2} & D^{(1)}_{0\ 0}(\chi)/\sqrt{2}\cr
0           & D^{(1)}_{-1\ 1-}(\chi)         &
D^{(1)}_{-1\ 1+}(\chi)         & D^{(1)}_{-1\ 0}(\chi)
\end{array}\right)~,
\end{equation}
which is the product of the $SU(4)$ matrix of spatial rotations
\begin{equation}
\widetilde{R}(\chi)=\exp(-i\alpha \widehat{S}_z^{(4)}) \exp(-i\beta \widehat{S}_y^{(4)})
\exp(-i\gamma \widehat{S}_z^{(4)})~,
\end{equation}
with the \( SO(3)\) generators $S_i^{(4)}$ defined in (\ref{S4a}), and the special  \( U(4)\) matrix
(${\rm det}\,\widetilde{U}(\chi)=-i$)
\begin{equation}
\label{U_0}
\widetilde{U}_0=\frac{1}{\sqrt{2}}
\left(\begin{array}{cccc}
0  &  1  &  -i  &  0  \cr
1  &  0  & 0  &  -1  \cr
-1  & 0  &  0  &  -1  \cr
0  &  -1  &  -i  &  0
\end{array}\right)~.
\end{equation}
Hence the unconstrained spin-0 and spin-2 gluon fields read (using Wigner D-functions)
\begin{eqnarray}
S^{\!(0)} =\left(\phi_1+\phi_2+\phi_3\right)/\sqrt{3}~,
\quad\quad\quad
S^{\!(2)}_{A} =
\sqrt{\frac{2}{3}}\left[
\left(\phi_3-\frac{1}{2}\left(\phi_1+\phi_2\right)\right)D^{(2)}_{A0}(\chi)
+\frac{\sqrt{3}}{2}\left(\phi_1-\phi_2\right)
D^{(2)}_{A2+}(\chi)
\right]~,
\end{eqnarray}
in terms of the principle axes variables, and the unconstrained spin-0 and spin-1 quark fields
\begin{eqnarray}
\psi_L^{\prime (0)} = \widetilde{u}^{(0)}_L~,&\quad\quad &
\psi_{L,i}^{\prime (1)}
= R_{ij}(\chi)\,\widetilde{u}^{(j)}_L~,
\nonumber\\
\psi_R^{\prime (0)} = \widetilde{v}^{(0)}_R~,&\quad\quad &
\psi_{R,i}^{\prime (1)}
= R_{ij}(\chi)\,\widetilde{v}^{(j)}_R~,
\end{eqnarray}
in terms of the intrinsic quark variables
\begin{equation}
\widetilde{u}_L
=(\widetilde{u}^{(0)}_L,\ \widetilde{u}^{(1)}_L,\ \widetilde{u}^{(2)}_L,\ \widetilde{u}^{(3)}_L)~,
\quad\quad
\widetilde{v}_R
=(\widetilde{v}^{(0)}_R,\ \widetilde{v}^{(1)}_R,\ \widetilde{v}^{(2)}_R,\ \widetilde{v}^{(3)}_R)~,
\end{equation}
satisfying anti-commutation relations
\begin{equation}
\{\widetilde{u}^{(\mu)}_L,\widetilde{u}^{(\nu)\dag}_L\}=\delta^{\mu\nu}~,
\quad\quad
\{\widetilde{v}^{(\mu)}_R,\widetilde{v}^{(\nu)\dag}_R\}=\delta^{\mu\nu}~.
\quad\quad
\{\widetilde{u}^{(\mu)}_L,\widetilde{v}^{(\nu)\dag}_R\}=0~.
\end{equation}

\subsection{\bf Unconstrained Hamiltonian in terms of scalar and rotational variables}

The transformation (\ref{patransfferm})-(\ref{U_0}) on the quark fields yields
(see the Appendix for details)
\begin{equation}
J^Q_i =R_{ij}(\chi)\ \widetilde{J}^Q_j~,
\quad\quad\quad\quad
K^Q_i=R_{ij}(\chi)\ \widetilde{K}^Q_j~,
\end{equation}
with the operators
\begin{eqnarray}
\label{JKtilde}
\widetilde{J}^Q_i:=-i\ \!\epsilon_{ijk}\left(\widetilde{u}^{(j)\dag}_L\widetilde{u}^{(k)}_L+
                         \widetilde{v}^{(j)\dag}_R\widetilde{v}^{(k)}_R\right)~,\quad\quad
\widetilde{K}^Q_i:=i\left(\widetilde{u}^{(0)\dag}_L\widetilde{u}^{(i)}_L+
             \widetilde{u}^{(i)\dag}_L\widetilde{u}^{(0)}_L\right)+
     i\left(\widetilde{v}^{(0)\dag}_R\widetilde{v}^{(i)}_R
             +\widetilde{v}^{(i)\dag}_R\widetilde{v}^{(0)}_R\right)~,
\end{eqnarray}
satisfying the same Lie-algebra (\ref{JKalgebra}) as the original $J^Q_i$ and $K^Q_i$.
Furthermore one finds (see App. for details)  that the minimal coupling part of the Hamiltonian
is diagonalised,
\begin{eqnarray}
g \psi^{\prime \dagger}\alpha_i S_{ij} {1\over 2}\tau_j \psi^{\prime} &=&
{g\over 2}(\phi_1+\phi_2+\phi_3)
\left(\widetilde{N}^{(0)}_L-\widetilde{N}^{(0)}_R\right)
+{g\over 2}
\sum_{ijk}^{\rm cyclic}\!\! \Bigg[\!\left(\phi_i-(\phi_j+\phi_k)\right)\!
\left(\widetilde{N}^{(i)}_L-\widetilde{N}^{(i)}_R\right)
\Bigg]~.
\end{eqnarray}
with the quark-number operators
\begin{equation}
\label{Ntilde}
\widetilde{N}^{(\mu)}_L=\widetilde{u}^{(\mu)\dag}_L\widetilde{u}^{(\mu)}_L~,\quad\quad\quad
\widetilde{N}^{(\mu)}_R=\widetilde{v}^{(\mu)\dag}_R\widetilde{v}^{(\mu)}_R~,\quad\quad\quad
\mu=0,1,2,3~.
\end{equation}
The transformation $\widetilde{U}$ in (\ref{U}) therefore leads to a diagonalisation
of the Dirac-Hamiltonian
for zero-momentum quarks in the background of a zero-momentum symmetric tensor field $S$.
The eigenvectors are the quark states $\widetilde{u}$ and $\widetilde{v}$
and the corresponding eigenvalues simple linear combinations of the eigenvalues of
the reduced gluon field.

Finally, the momenta canonically conjugate to the $S^{\!(0)}$ and $S^{\!(2)}_{A}$
are
\begin{eqnarray}
\label{new-mom1}
-i{\partial\over\partial S^{\!(0)}} &=&-i\left(
{\partial\over\partial \phi_{1}}+{\partial\over\partial \phi_{2}}+{\partial\over\partial \phi_{3}}\right)/\sqrt{3}~,\\
\label{new-mom2}
-i{\partial\over\partial S^{\!(2)}_{A}} &=&
\sqrt{\frac{2}{3}}\left[
-i\left({\partial\over\partial \phi_{3}}-\frac{1}{2}\left({\partial\over\partial \phi_{1}}
+{\partial\over\partial \phi_{2}}\right)
\right)D^{(2)}_{A 0}(\chi)
-\frac{\sqrt{3}}{2}i\left({\partial\over\partial \phi_{1}}-{\partial\over\partial \phi_{2}}\right)
D^{(2)}_{A 2+}(\chi)\right]\nonumber\\
&&+\frac{1}{\sqrt{2}}
       \Bigg[ D^{(2)}_{A 1+}(\chi) \frac{\xi^G_1}{\phi_2 - \phi_3}
             +D^{(2)}_{A 1-}(\chi) \frac{\xi^G_2}{\phi_3 - \phi_1}
             +D^{(2)}_{A 2-}(\chi) \frac{\xi^G_3}{\phi_1 - \phi_2}
       \Bigg]~,
\end{eqnarray}
using the intrinsic angular momenta
\begin{eqnarray}
\xi^G_i:=
\xi_i-\widetilde{J}^Q_i~,\quad\quad i=1,2,3
~,\quad\quad
[\xi^G_i,\xi^G_j]=-i\epsilon_{ijk}\xi^G_k~,
\end{eqnarray}
where
\begin{equation}
\xi_i:=-i{\cal M}^{-1}_{ij}{\partial\over\partial \chi_{i}}~,
\quad
{\cal M}_{ij}:=-{1\over 2}\ \epsilon_{jst}\!
        \left(R^T{\partial R\over\partial \chi_{i}} \right)_{st}~,
\quad
[\xi_i,\xi_j]=-i\epsilon_{ijk}\xi_k~.
\end{equation}
For the case of Euler angles $\chi=(\alpha,\beta,\gamma)$ we have
\begin{eqnarray}
{\cal M}^{-1}=
\left(\begin{array}{ccc}
\sin\gamma & -\cos\gamma / \sin\beta  & \cos\gamma \cot\beta \cr
\cos\gamma &  \sin\gamma / \sin\beta  & -\sin\gamma \cot\beta  \cr
0&0&1
\end{array}\right)~.
\nonumber
\end{eqnarray}
Hence we obtain
\begin{eqnarray}
\label{HphysA}
 &&\!\!\!\!\!\!\!\!\!\!\!\!\!\!\!\!\!\!\!\!\!\!\!\!\!\!\!\!\!\!\!\!
 {1\over 2}\sum_{m,n} \Bigg[
-\left(\frac{\partial}{\partial S_{mn}}\right)^2
-\left[\gamma^{-1}_{mn}(S)-\delta_{mn}\mbox{tr}(\gamma^{-1}(S))\right]
    \frac{\partial}{\partial S}_{mn}
+B_{mn}^2(S)\Bigg]=
\nonumber\\
\!\!&=&\!\!{1\over 2}\!\sum^{\rm
cyclic}_{ijk}\!\! \Bigg[\!\!-\left({\partial\over\partial \phi_i}\right)^2 -{2\over \phi_i^2-
\phi_j^2}\!\left(\phi_i{\partial\over\partial
\phi_i}-\phi_j{\partial\over\partial \phi_j}\right)
 +{1\over 2} \left({\xi^G_i\over \phi_j-\phi_k}\right)^2
+\!  g^2 \phi_j^2 \phi_k^2\Bigg]
\end{eqnarray}
and
\begin{equation}
J^G_i := -2i\epsilon_{ijk}S_{aj}{\partial\over\partial S_{ak}}=R_{ij}(\chi)\ \xi^G_j~.
\end{equation}

Altogether, after rescaling the fields
$\phi_i \rightarrow  g^{-1/3} \phi_i~ (i=1,2,3)$,
and then reinstalling $g^2\rightarrow g^2/V$, we obtain the unconstrained Hamiltonian
of $SU(2)$ Dirac-Yang-Mills quantum mechanics in the final form
\begin{equation}
\label{Hphys}
H={g^{2/3}\over V^{1/3}}\Bigg[{\cal H}^{G}+{\cal H}^{D}+{\cal H}^{C}\Bigg]+H_{m}~,
\end{equation}
with
\begin{eqnarray}
\label{HphysG}
{\cal H}^{G}&:=&{1\over 2}\sum^{\rm
cyclic}_{ijk}\!\! \Bigg(\!\!-{\partial^2\over\partial \phi_i^2} -{2\over \phi_i^2-
\phi_j^2}\!\left(\phi_i{\partial\over\partial
\phi_i}-\phi_j{\partial\over\partial \phi_j}\right)
 +(\xi_i-\widetilde{J}^Q_i)^2 \!\! {\phi_j^2+\phi_k^2\over (\phi_j^2-\phi_k^2)^2}
+\!  \phi_j^2 \phi_k^2\Bigg)~,
\\
\label{HphysD}
{\cal H}^{D}&:=&\!\! {1\over 2}(\phi_1+\phi_2+\phi_3)\!
\left(\widetilde{N}^{(0)}_L-\widetilde{N}^{(0)}_R\right)
\!+\!
{1\over 2}\sum^{\rm cyclic}_{ijk}\!\! (\phi_i-(\phi_j+\phi_k))\!
\left(\widetilde{N}^{(i)}_L-\widetilde{N}^{(i)}_R\right)~,
\\
\label{HphysC}
{\cal H}^{C}&:=&
\!-{1\over 4}\! \sum^{\rm cyclic}_{ijk}\!\!
      {(\xi_i-\widetilde{J}^Q_i)^2-(\xi_i-i\widetilde{K}^Q_i)^2 \over (\phi_j+\phi_k)^2}~,
\end{eqnarray}
the quark-mass term
\begin{eqnarray}
\label{Hphysm}
H_{m}&:=&{1\over 2} m\! \left[\!\left(\widetilde{u}^{(0)\dag}_L\widetilde{v}^{(0)}_R\!
         +\sum_{i=1}^3\widetilde{u}^{(i)\dag}_L\widetilde{v}^{(i)}_R\right)\! +h.c.\right]~,
\end{eqnarray}
and the total angular momentum
\begin{equation}
\label{Jphys}
J_i=R_{ij}(\chi)\left(\xi^G_j
+\widetilde{J}^Q_j\right)
=R_{ij}(\chi)\ \xi_j~.
\end{equation}
Since the Jacobian of  (\ref{patransf}) is
$|\partial S/\partial(\alpha,\beta,\gamma,\phi)| \propto
\sin\beta \prod_{i<j}\left(\phi_i- \phi_j\right)$,
the matrix elements of an operator $O$ are given as
\begin{equation}
\langle\Phi'|O|\Phi\rangle\! \propto\!\!
\int\!\! d \overline{\widetilde{u}}_L d \widetilde{u}_L d \overline{\widetilde{v}}_R d \widetilde{v}_R
\int\!\! d\alpha \sin\beta d\beta d\gamma\!\!\!\!\!\!\!\!\!\!\!\!\!
\int\limits_{0<\phi_1<\phi_2<\phi_3}\!\!\!\!\!\!\!\!\!\!\!\!\! \Big[\!\!\prod^{\rm cyclic}\! d\phi_i
\! \left(\phi_j^2- \phi_k^2\right)\Big]
\Phi'^* \, O\, \Phi.
\label{measure}
\end{equation}
The representation of the unconstrained Hamiltonian (\ref{Hphys})-(\ref{Hphysm}) and total
angular momentum (\ref{Jphys}) of $SU(2)$ Dirac-Yang-Mills quantum mechanics of spatially
constant quark and gluon fields, is the main result of the present work and is new to the best
of my knowledge. In order to keep the formulas simple, only the 1-flavor case
has been shown in this work. The generalization of the Hamiltonian (\ref{Hphys})-(\ref{Hphysm}) to flavor-numbers
larger than one is trivial: The operators $\widetilde{N}^{(\mu)}_{L(R)},\widetilde{J}^{Q}_{i},\widetilde{K}^{Q}_{i}$
and $H_{m}$ become sums over different flavors.

\subsection{\bf Symmetries of the Hamiltonian}

Due to $[J_i,\xi_j]=0$, the angular momenta $J_i$ commute with the Hamiltonian,
and the eigenstates of $H$ can be characterized by the quantum numbers $J$ and $M$ of total spin
of the quark-gluon system.
Furthermore $H$ is invariant under cyclic permutations $\sigma_{123}$ of the three indices $1,2,3$,
parity transformations $P$
\begin{eqnarray}
\label{parity}
P:\quad\quad (\phi_i\rightarrow -\phi_i\ \ (i=1,2,3))\quad \wedge \quad
 (\widetilde{u}^{(\mu)}_L\leftrightarrow \widetilde{v}^{(\mu)}_R\ \ (\mu=0,1,2,3))~,
\end{eqnarray}
time inversion $T$ (anti-unitary)
\begin{equation}
\label{time}
T:\quad\quad (\phi_i\rightarrow -\phi_i\ \ (i=1,2,3))\quad \wedge \quad
 (|0\rangle \rightarrow |8\rangle \quad \wedge \quad
 \widetilde{u}_L^{(\mu)}\rightarrow \widetilde{u}_L^{(\mu)\dag},
 \widetilde{v}_R^{(\mu)}\rightarrow -\widetilde{v}_R^{(\mu)\dag}
\ (\mu=0,1,2,3)) ~,
\end{equation}
where $|0\rangle$ and $|8\rangle$ denote the energy degenerate and complex conjugate\footnote{
The conditions $\widetilde{u}_L^{(\mu)}|0\rangle =0 \ \wedge\ \widetilde{v}_R^{(\mu)}|0\rangle =0$ transform
under $T$ into $\widetilde{u}_L^{(\mu)\dag}|0\rangle^* =0 \ \wedge\ \widetilde{v}_R^{(\mu)\dag}|0\rangle^* =0$,
from which $|0\rangle^* \equiv |8\rangle$ follows.
}
quark states
with no and with all levels filled, and charge conjugation $C$
\begin{eqnarray}
\label{charge}
C:\quad\quad
 (|0\rangle \rightarrow |8\rangle \quad \wedge \quad
 \widetilde{u}_L^{(0)}\rightarrow \widetilde{v}_R^{(0)\dag},
 \widetilde{v}_R^{(0)}\rightarrow -\widetilde{u}_L^{(0)\dag}\quad \wedge \quad
\widetilde{u}_L^{(i)}\rightarrow -\widetilde{v}_R^{(i)\dag},
\widetilde{v}_R^{(i)}\rightarrow \widetilde{u}_L^{(i)\dag}\ (i=1,2,3)~) ~,
\end{eqnarray}
such that
\begin{equation}
\label{symH1}
[H,J_i]=0~,\ \ \ \ \ \ [H,\sigma_{123}]=0~,\ \ \ \ \ \ [H,P]=0~,\ \ \ \ \ \ [H,T]=0~,\ \ \ \ \ \ [H,C]=0~.
\end{equation}
Furthermore, $H$ commutes with the total number $N$ of quarks
\begin{equation}
\label{symH2}
[H, N]=0~,\quad\quad N:=N_L+N_R~,\quad\quad
N_{L(R)}:=\sum_{\mu=0,1,2,3}\widetilde{N}^{(\mu)}_{L(R)}~.
\end{equation}
The eigenstates of $H$ can therefore be characterized by the total number $N=0,1,2,..,8$ of quarks.
Due to the charge-conjugation symmetry $C$ in (\ref{charge}), the states with quark numbers $N$ and $8-N$ are
degenerate in energy.

It is important to notice, that, in contrast to the parts ${\cal H}^{G}$ in (\ref{HphysG}) and
${\cal H}^{C}$ in (\ref{HphysC}), which both are invariant under the transformations
$\phi_i\rightarrow -\phi_i$ and $L\longleftrightarrow R$ separately, the part ${\cal H}^{D}$ in (\ref{HphysD})
is invariant under the combination $P$ in (\ref{parity}). This will be seen to be crucial for the
lowering of the ground state energy of the quark-gluon system in comparison to the pure-gluon case
and for the appearance of a quark-condensate.

\subsection{\bf Boundary conditions}

Note that for the pure-gluon case, $H$ reduces to
\begin{equation}
{\rm For}\ N=0:\quad\quad  H\big|_{N|\Phi\rangle=0}={\cal H}^{G}_{\widetilde{J}^Q=0}~.
\end{equation}
Its eigenstates
\begin{eqnarray}
\!\!\!\!\!\!\!\!\!\!\!\!
{\cal H}^G_{\widetilde{J}^Q=0} |\Phi^{\!(J)\pm}_{n,M}\rangle\!\!\!\!\!&=&\!\!\!\!\!
                                   \epsilon^{(J)\pm}_{n} |\Phi^{\!(J)\pm}_{n,M}\rangle~,
\end{eqnarray}
can be chosen to have definite angular momentum and parity quantum numbers,
\begin{eqnarray}
 |\Phi^{\!(J)\pm}_{n,M}\rangle  =
 \sum_{M^\prime} \Phi^{\!(J,M^\prime)\pm}_{n}(\phi_1,\phi_2,\phi_3)|J\, M\, M^\prime\rangle~,
\quad \quad
|J\, M\, M^\prime\rangle = i^J \sqrt{{J+1\over 8\pi^2}} D^{(J)}_{M M^\prime}(\chi)~,
\end{eqnarray}
and are known with high accuracy \cite{pavel1}.
The requirement of Hermiticity of ${\cal H}^G_{\widetilde{J}^Q=0}$
in the region bounded by the three boundary planes $\phi_1=0~,\phi_1=\phi_2~, \phi_2=\phi_3~$
and at positive infinity, leads to the conditions
\begin{equation}
\label{gluonbc}
\partial_{\phi_1}\Phi^{\!(J,M^\prime)+}_{n}\Big|_{\phi_1=0}= 0~,
\quad  \wedge  \quad
\Phi^{\!(J,M^\prime)-}_{n}\Big|_{\phi_1=0}= 0~,
\quad  \wedge  \quad
\Phi^{\!(J,M^\prime)\pm}_{n}\Big|_{\phi_2=\phi_1}= {\rm finite}
\quad  \wedge  \quad
\Phi^{\!(J,M^\prime)\pm}_{n}\Big|_{\phi_3=\phi_2}={\rm finite}~.
\end{equation}
Finally,
normalisability of the wave functions requires that the wave functions
vanish sufficiently fast at infinity.

In the general case of non-vanishing quark-number,
one can put the quark boundary condition (possible only for the case of half-filling $N=4$),
to be eigenstates under the $C$ symmetry (\ref{charge})\footnote{\label{fermbc2}
Other possibilities,  e.g.
\begin{equation}
{\rm For}\ N=2,4,6:\quad\quad
\widetilde{N}^{(\mu)}_L|\Psi\rangle=\widetilde{N}^{(\mu)}_R|\Psi\rangle~,\quad\quad \forall\mu=0,1,2,3~,
\nonumber
\end{equation}
would restrict the quark states to positive-parity states and hence lead to a vanishing ${\cal H}^D$ in (\ref{HphysD}).
}
\begin{equation}
\label{fermbc}
{\rm For}\ N=4:\quad\quad
\widetilde{N}^{(\mu)}_L|\Psi\rangle=\left(1-\widetilde{N}^{(\mu)}_R\right)|\Psi\rangle~,\quad\quad \forall\mu=0,1,2,3~,
\end{equation}
invariant under all symmetries (\ref{symH1}) and (\ref{symH2}) of the Hamiltonian (\ref{Hphys}).

From all quark states $\Psi$ satisfying the boundary condition (\ref{fermbc})
 one can build the quark spin  and parity eigenstates
\begin{equation}
({\mathbf{J}}^{Q})^2|\Psi^{\!(J)\pm}_{M}\rangle\!\!=\!J(J+1)|\Psi^{\!(J)\pm}_{M}\rangle~,
\quad \ J^{Q}_z|\Psi^{\!(J)\pm}_{M}\rangle\!\!=\!M|\Psi^{\!(J)\pm}_{M}\rangle~,
\quad \ P|\Psi^{\!(J)\pm}_{M}\rangle\!\!=\!\pm|\Psi^{\!(J)\pm}_{M}\rangle~,
\end{equation}
and diagonalize ${\cal H}$ in the basis
\begin{equation}
|J_{1\, n}^{P_1} \otimes J_2^{P_2}\rangle^{\!(J)P}_{M}
=\delta_{P_1\cdot P_2\,-\, P}\sum_{M_1,M_2}
C^{J\, M}_{J_1\, M_1\, J_2\, M_2}\
|\Phi^{\!(J_1)P_1}_{n,M_1}\rangle\otimes |\Psi^{\!(J_2)P_2}_{M_2}\rangle~.
\end{equation}

\section{\large\bf Calculation of the spin-0 energy spectrum}

For total spin-0 (${\mathbf{J}^2}=\xi_1^2+\xi_2^2+\xi_3^2=0$) and for massless quarks
the Hamiltonian (\ref{Hphys}) reduces to
\begin{eqnarray}
\label{Hphysm0}
H_{0}={g^{2/3}\over V^{1/3}}\Bigg[{\cal H}^{G}_{\xi=0}+{\cal H}^{D}+{\cal H}^{C}_{\xi=0}\Bigg]
\end{eqnarray}
In addition to the above stated symmetries the $m=0$ Hamiltonian (\ref{Hphysm0})
and the boundary conditions (\ref{fermbc}) are invariant under independent global $U(1)$ phase rotations
of the left- and the right-handed quarks, and the total numbers $N_{L(R)}$ of left-(right-) handed
quarks are good quantum numbers.

Let us first consider 4-quark states with 2 left-handed and 2 right-handed quarks
with the boundary condition (\ref{fermbc}), satisfied by the 6 states
\begin{eqnarray}
\label{Psipm}
|\Psi_{i}^{\pm}\rangle :=\frac{1}{\sqrt{2}}\left(
               \widetilde{u}^{(j)\dag}_L\widetilde{u}^{(k)\dag}_L
               \widetilde{v}^{(0)\dag}_R\widetilde{v}^{(i)\dag}_R|0\rangle
               \ \pm\
               \widetilde{u}^{(0)\dag}_L\widetilde{u}^{(i)\dag}_L
               \widetilde{v}^{(j)\dag}_R\widetilde{v}^{(k)\dag}_R|0\rangle
\right)
\quad\quad\quad\quad
i,j,k=1,2,3 \quad {\rm cycl.}~,
\end{eqnarray}
which are $\pm$ parity-eigenstates, and construct the
$({\mathbf{J}}^Q)^2=(\widetilde{\mathbf{J}}^Q)^2$ eigenstates
\begin{eqnarray}
|\Psi^{(0)\pm}\rangle &:=&
\left(|\Psi_{1}^{\pm}\rangle+|\Psi_{2}^{\pm}\rangle+|\Psi_{3}^{\pm}\rangle\right)/\sqrt{3}~,
\nonumber\\
|\Psi^{(2)\pm}_A\rangle &:=&D^{(2)}_{A0}(\chi)|\Psi^{(2,0)\pm}\rangle
                           +  D^{(2)}_{A2+}(\chi)|\Psi^{(2,2+)\pm}\rangle~,
\end{eqnarray}
with
\begin{eqnarray}
|\Psi^{(2,0)\pm}\rangle &:=&\sqrt{\frac{2}{3}}
 \left(|\Psi^{\pm}_3\rangle-\frac{1}{2}\left(|\Psi^{|\pm}_1\rangle+|\Psi^{\pm}_2\rangle\right)\right),
\quad
|\Psi^{(2,2+)\pm}\rangle :=\frac{1}{\sqrt{2}}\left(|\Psi^{\pm}_1\rangle-|\Psi^{\pm}_2\rangle\right)~.
\end{eqnarray}
Furthermore, let
\begin{eqnarray}
|\Phi^{\!(0)\pm}_{n}\rangle &=& \Phi^{\!(0)\pm}_{n}(\phi)|0\, 0\, 0\rangle~,\quad n=0,1,2,3,..~,
\nonumber\\
|\Phi^{\!(2)\pm}_{m,A}\rangle &=&
\Phi^{(2,0)\pm}_{m}(\phi)|2\, A\, 0\rangle
                        +\Phi^{(2,2+)\pm}_{m}(\phi) |2\, A\, 2+\rangle
~,\quad m=0,1,2,3,..~,
\end{eqnarray}
be the complete set of $\pm$-parity spin-0 and spin-2
eigenstates \cite{pavel1}  of the pure-gluon part
${\cal H}^G_{\widetilde{J}^Q=0}$ of the Hamiltonian (\ref{Hphys}).
Since the action of the $(\widetilde{J}_i^Q)^2$ on the $\Psi^{(0)}$,\, $\Psi^{(2,0)}$ and $\Psi^{(2,2+)}$,
is the same as that
of the $\xi_i^2$ on the Wigner D-functions $D^{(0)}_{00}=1$,\, $D^{(2)}_{A0}$ and $D^{(2)}_{A2+}$,
respectively, the positive-parity spin-0 combinations
\begin{eqnarray}
|0_n^\pm\otimes 0^\pm\rangle^{(0)+} &=&
\Phi^{\!(0)\pm}_{n}(\phi) |\Psi^{\!(0)\pm}\rangle~,
\quad\quad\quad\quad\quad\quad\quad\ n=0,1,2,3,...
\nonumber
\\
|2_m^\pm\otimes 2^\pm\rangle^{(0)+} &=&
\Phi^{\!(2,0)\pm}_{m}(\phi)|\Psi^{\!(2,0)\pm}\rangle+\Phi^{\!(2,2+)\pm}_{m}(\phi)|\Psi^{\!(2,2+)\pm}\rangle~,
\quad m=0,1,2,3,...
\label{repr}
\end{eqnarray}
which are eigenstates of $C$, $P$ and $T$ symmetries (\ref{parity})-(\ref{charge}),
form a the complete set of eigenfunctions of ${\cal H}^{G}_{\xi=0}$,
\begin{eqnarray}
{\cal H}^{G}_{\xi=0}&&=\sum_n  \Big( \epsilon^{(0)+}_{n}\
                                   |0_n^+\otimes 0^+\rangle^{(0)+} \langle 0_n^+\otimes 0^+|^{(0)+}\  +\
                                     \epsilon^{(0)-}_{n}\
                                   |0_n^-\otimes 0^-\rangle^{(0)+} \langle 0_n^-\otimes 0^-|^{(0)+}\Big)
\nonumber\\
              &&+ \sum_m  \Big( \epsilon^{(2)+}_{m}\
                                   |2_m^+\otimes 2^+\rangle^{(0)+} \langle 2_m^+\otimes 2^+|^{(0)+}\
                               +\ \epsilon^{(2)-}_{m}\
                                   |2_m^-\otimes 2^-\rangle^{(0)+} \langle 2_m^-\otimes 2^-|^{(0)+}\Big)~.
\end{eqnarray}
The interactions  ${\cal H}^D$ and  ${\cal H}^C_{\xi=0}$ can be written in the representation
(\ref{repr}) as
\begin{eqnarray}
\!\!\!{\cal H}^D\!\!\!&=&\!\!\!
             {2\over \sqrt{3}}\sum_{n,n^\prime}
\langle\Phi^{\!(0)+}_{n^\prime}||\!\left(\!\phi_3\right)^{\!(0)}\!\!||\Phi^{\!(0)-}_{n}\rangle
                   \Big(|0_{n^\prime}^+\otimes 0^+\rangle^{(0)+} \langle 0_{n}^-\otimes 0^-|^{(0)+}+\ h.c.\Big)
\nonumber\\
           &&\!\!\!\!\!\!
             -{2\over \sqrt{3}}\sum_{m,n}\Big[
\langle\Phi^{\!(0)+}_{n}||\!\left(\!\phi_3\right)^{\!(2)}\!\!||\Phi^{\!(2)-}_{m}\rangle
                   \Big( |0_{n}^+\otimes 0^+\rangle^{(0)+} \langle 2_m^-\otimes 2^-|^{(0)+}+\ h.c.\Big)
\nonumber\\
&&\quad\quad\quad\quad\quad\quad\quad\quad
+\langle\Phi^{\!(0)-}_{n}||\!\left(\!\phi_3\right)^{\!(2)}\!\!||\Phi^{\!(2)+}_{m}\rangle
                   \Big( |0^-_n\otimes 0^-\rangle^{(0)+} \langle 2^+_m\otimes 2^+|^{(0)+}
+\ h.c.\Big)\Big]
\nonumber\\
           &&\!\!\!\!\!\!
            +{2\over \sqrt{15}}\sum_{m,m^\prime}
\Big(\langle\Phi^{\!(2)+}_{m^\prime}||\!\left(\!\phi_3\right)^{\!(0)}\!\!||\Phi^{\!(2)-}_{m}\rangle
-{\sqrt{7}\over 2}\langle\Phi^{\!(2)+}_{m^\prime}||\!\left(\!\phi_3\right)^{\!(2)}\!\!||\Phi^{\!(2)-}_{m}\rangle\Big)
                   \Big( |2_{m^\prime}^+\otimes 2^+\rangle^{(0)+} \langle 2_m^-\otimes 2^-|^{(0)+}+\ h.c.\Big)
\end{eqnarray}
and
\begin{eqnarray}
\!\!\!{\cal H}^C_{\xi=0}\!\!\!&=&\!\!\!
             {2\over \sqrt{3}}\sum_{n,n^\prime}\
\langle\Phi^{\!(0)-}_{n^\prime}||\!\!\left({1\over (\phi_1+\phi_2)^2}
\right)^{\!\!\!(0)}\!\!\!\!||\Phi^{\!\!(0)-}_{n}\rangle\
                   |0_{n^\prime}^{-}\otimes 0^-\rangle^{(0)+} \langle 0_{n}^{-}\otimes 0^-|^{(0)+}
\nonumber\\
           &&\!\!\!\!\!\!
             +{1\over \sqrt{3}}\sum_{m,n}\
\langle\Phi^{\!(0)-}_{n}||\!\!\left({1\over (\phi_1+\phi_2)^2}
\right)^{\!\!\!(2)}\!\!\!\!||\Phi^{\!(2)-}_{m}\rangle\
                   \Big( |0_{n}^-\otimes 0^-\rangle^{(0)+} \langle 2_m^-\otimes 2^-|^{(0)+}+\ h.c.\Big)
\nonumber\\
           &&\!\!\!\!\!\!
            -{1\over \sqrt{15}}\!\!\sum_{m,m^\prime}\!\!\!
            \Big(
\langle\Phi^{\!(2)-}_{m^\prime}||\!\!
\left({1\over (\phi_1+\phi_2)^2}\right)^{\!\!\!(0)}\!\!\!\!||\Phi^{\!(2)-}_{m}\rangle
\nonumber\\
&&\quad\quad\quad\quad\quad\quad\quad\quad
 + \sqrt{7}\langle\Phi^{\!(2)-}_{m^\prime}||\!
 \left({1\over (\phi_1+\phi_2)^2}\right)^{\!\!\!(2)}\!\!\!\!||\Phi^{\!(2)-}_{m}\rangle
            \Big)\
                    |2_{m^\prime}^{-}\otimes 2^-\rangle^{(0)+} \langle 2_{m}^{-}\otimes 2^-|^{(0)+}
~,
\end{eqnarray}
with the abbreviations
\begin{eqnarray}
\left(\!\phi_3\right)^{\!(0)}:&=&\left(\phi_1+\phi_2+\phi_3\right)/\sqrt{3}~,
\nonumber\\
\left(\!\phi_3\right)^{\!(2)}_{A}:&=&
\sqrt{\frac{2}{3}}
\left(\phi_3-{1 \over 2}(\phi_1+\phi_2)\right) D^{(2)}_{A0}(\chi)
+\frac{1}{\sqrt{2}}\left(\phi_1-\phi_2\right) D^{(2)}_{A2+}(\chi)~.
\nonumber
\end{eqnarray}

Truncating the space of states at 30 nodes and diagonalising the total $H_0$ one obtains the
energy spectrum shown in the third column of Tab.1 and the second spectrum in Fig.1.
The 3 numbers in brackets behind the energy values in the table give the contributions
from the parts ${\cal H}^G_{\xi=0}$, ${\cal H}^D$, and ${\cal H}^C_{\xi=0}$ separately.
The second column of Tab.1 and the first spectrum in Fig.1 show the energy levels
for the corresponding pure-gluon case.
Since the gluonic matrix elements are calculated with high accuracy, the errors are expected
to be smaller than the last digits shown and lie inside the lines.
The lowest state is (up to contributions $\leq 0.1$)\footnote{
The coefficients in (\ref{groundstate}) satisfy
$\sum_n\left[(c_n^{(0)+})^2+(c_n^{(0)-})^2\right]+\sum_m\left[(c_m^{(2)+})^2+(c_m^{(2)-})^2\right]=1~.$
}
\begin{eqnarray}
\label{groundstate}
|0\rangle^{(0)+}_{QG(2L2R)}\!\!\!\!\!\!
&=&\!\!\!\!\!\sum_n\left[c_n^{(0)+}|0_n^+\!\otimes 0^+\rangle^{(0)+}
              +c_n^{(0)-}|0_n^-\!\otimes 0^-\rangle^{(0)+}\right]
   +\sum_m\left[c_m^{(2)+}|2_m^+\!\otimes 2^+\rangle^{(0)+}
               +c_m^{(2)-}|2_m^-\!\otimes 2^-\rangle^{(0)+}\right]\nonumber\\
\!\!\!\! = 0.80 \!\!\!\!\! &&\!\!\!\!\!\!\!\!\!\!\! |0_0^+\!\otimes\! 0^+\rangle^{(0)+}\!\!
                      -0.38 |2_0^+\!\otimes\! 2^+\rangle^{(0)+}\!\!
                      -0.28  |0_0^-\!\otimes\! 0^-\rangle^{(0)+}\!\!
                      -0.24 |2_0^-\!\otimes\! 2^-\rangle^{(0)+}\!\!
+0.19 |0_1^+\!\otimes\! 0^+\rangle^{(0)+}\!\!
                      +\! ..,
\end{eqnarray}
and its energy $E^{(0)+}_{0\ QG(2L2R)}=3.22\, g^{2/3}/V^{1/3}$ about $20\%$ lower than the lowest energy
$E^{(0)+}_{0\ G}=4.117\, g^{2/3}/V^{1/3}$ for the pure-gluon case.
It is the lowest of all spin-0 states for different numbers of massless reduced quarks and
therefore constitutes the ground state.
Note that responsible for the lowering of the energy in the presence of quarks is ${\cal H}^D$,
which leads to transitions between gluon states with positive and negative parity,
of course accompanied by a corresponding transition between quark states with opposite parity in order to be
invariant under the total parity transformation $P$ in (\ref{parity}).
At the same time, ${\cal H}^D$ is responsible for the formation of a quark condensate.
The quark operator
\begin{eqnarray}
O_{2L2R}:=\sum_{i,j,k}^{\rm cyclic}\left[(\widetilde{u}^{(0)\dag}_L\widetilde{v}^{(0)}_R)
                     (\widetilde{u}^{(i)\dag}_L\widetilde{v}^{(i)}_R)
                     (\widetilde{v}^{(j)\dag}_L\widetilde{u}^{(j)}_R)
                     (\widetilde{v}^{(k)\dag}_L\widetilde{u}^{(k)}_R)+\ h.c.\right]
\end{eqnarray}
which connects the parity-doublet partners in the quark wave function (\ref{Psipm}),
has non-vanishing expectation values.
In particular for the lowest state (\ref{groundstate}), we find the quark condensate
according to the simple formula\begin{eqnarray}
\langle 0|(\overline{\psi}\,\psi)^4|0\rangle^{(0)+}_{QG(2L2R)}
=24\Big(\sum_n\left[(c_n^{(0)+})^2-(c_n^{(0)-})^2\right]+
           \sum_m\left[(c_m^{(2)+})^2-(c_m^{(2)-})^2\right]\Big)\simeq 16~,
\end{eqnarray}
in addition to a gluon condensate (to be calculated) as in the pure-gluon case.
Note however, that\\ $\langle 0|\overline{\psi} \psi|0\rangle^{\! (0)+}_{QG(2L2R)}\!=\! 0$.

For the case of 4-quark states with 3 left-handed and 1 right-handed quark or vice versa
with the boundary condition (\ref{fermbc}), we have the 8 states\footnote{
Due to the possibility of independent global $U(1)$ phase rotations of the left- and the
right-handed quark fields for $m=0$, each state is in fact a continous family of energy-degenerate
states. This is not the case
for the states (\ref{Psipm}) of equal numbers of 2 left- and 2 right- handed quarks,
in particular for the ground state (\ref{groundstate}), which are non-degenerate.
}
\begin{eqnarray}
|\widetilde{\Psi}_{0}^{\pm}\rangle &:=&\frac{1}{\sqrt{2}}\left(
               \widetilde{u}^{(1)\dag}_L\widetilde{u}^{(2)\dag}_L\widetilde{u}^{(3)\dag}_L
               \widetilde{v}^{(0)\dag}_R|0\rangle
               \ \pm\
               \widetilde{v}^{(1)\dag}_R
               \widetilde{v}^{(2)\dag}_R\widetilde{v}^{(3)\dag}_R
               \widetilde{u}^{(0)\dag}_L|0\rangle
\right)
\nonumber\\
|\widetilde{\Psi}_{i}^{\pm}\rangle &:=&\frac{1}{\sqrt{2}}\left(
               \widetilde{u}^{(0)\dag}_L\widetilde{u}^{(j)\dag}_L\widetilde{u}^{(k)\dag}_L
               \widetilde{v}^{(i)\dag}_R|0\rangle
               \ \pm\
               \widetilde{v}^{(0)\dag}_R
               \widetilde{v}^{(j)\dag}_R\widetilde{v}^{(k)\dag}_R\widetilde{u}^{(i)\dag}_L|0\rangle
\right)
\quad\quad\quad\quad
i,j,k=1,2,3 \quad {\rm cycl.}~.
\end{eqnarray}
A similar calculation as for the first case yields the
energy spectrum shown in the forth column of Tab.1 and the third spectrum of Fig.1.
The energy $E^{(0)+}_{0\ QG(3L1R/1L3R)}=3.42\, g^{2/3}/V^{1/3}$ of its lowest state is also
considerably lower than the ground state energy for the pure-gluon case
but a little higher than for the first case of two right- and two left-handed quarks.
It  could be identified with the sigma-antisigma excitation in this massless 1-flavor 2-color
investigation, its mass would be about one fifth of the first glueball excitation and
therefore of the right order of magnitude.

Finally we mention, that for the case of 4 right-handed or 4 left-handed quarks
with the boundary condition (\ref{fermbc}), we have the 2 states
\begin{eqnarray}
|\widetilde{\widetilde{\Psi}}_{0}^{\pm}\rangle &:=&\frac{1}{\sqrt{2}}\left(
               \widetilde{u}^{(1)\dag}_L\widetilde{u}^{(2)\dag}_L\widetilde{u}^{(3)\dag}_L
               \widetilde{u}^{(0)\dag}_L|0\rangle
               \ \pm\
               \widetilde{v}^{(1)\dag}_R
               \widetilde{v}^{(2)\dag}_R\widetilde{v}^{(3)\dag}_R
               \widetilde{v}^{(0)\dag}_R|0\rangle
\right)~,
\end{eqnarray}
leading to an energy spectrum which
is the sum-set of the energy levels of the positive- and negative-parity
pure-gluon cases, $\{E^{(0)+}_{i\ QG(4L/4R)}\}
=\{E^{(0)+}_{n\ G}\}\bigcup \{E^{(0)-}_{m\ G}\}$,
shown in the last column of Tab.1 and the last spectrum in Fig.1 (note that the first three
and the 5th energy value coincide with $E^{(0)+}_{i\ G}, i=0,1,2,3$, whereas the 4th energy
level is the lowest $E^{(0)-}_{0\ G}$ level).

All other number of particles and boundary conditions other than (\ref{fermbc})
do not lower the energy in comparison with the pure gluon case.
Note, that if one imposed the boundary condition of Footnote \ref{fermbc2} instead of (\ref{fermbc})
on the 4 quarks, ${\cal H}^D$ made
no contribution and one would obtain a ground-state energy higher than that for the pure-gluon case.
Also for boundary conditions which are rotationally invariant combinations of (\ref{fermbc}) and that in
Footnote \ref{fermbc2},
putting one condition on the reduced spin-0 quarks and the other on the reduced spin-1 quarks,
the energy is not lowered.

We have considered here the simplest case of one quark-flavor with mass $m=0$.
A non-vanishing quark-mass will lead to transitions between the states of the three massless cases discussed here.
The generalization to $m>0$, higher total spins and two or three flavors is straightforward and will be the subject
of future work.

\begin{table}
\centering
$\begin{array}{|c|c|c|c|c|}
i & E^{(0)+}_{i\ G}\ [g^{2/3}/V^{1/3}] & E^{(0)+}_{i\ QG(2L2R)}\ [g^{2/3}/V^{1/3}]
& E^{(0)+}_{i\ QG(3L1R/1L3R)}\ [g^{2/3}/V^{1/3}]& E^{(0)+}_{i\ QG(4L/4R)}\ [g^{2/3}/V^{1/3}] \cr
\hline
     0 & 4.117 &  3.22\ \  (\hspace*{2mm} 5.63,-2.43,0.02)&  3.42\ \  (\hspace*{2mm} 5.03,-1.82,0.21) & 4.117\ \ (4.117,0,0)\cr
     1 & 6.386 &   4.31\ \  (\hspace*{2mm} 7.43,-3.13,0.01)&  4.28\ \  (\hspace*{2mm} 5.03,-1.28,0.53)& 6.386\ \ (6.386,0,0)\cr
     2 & 7.973 &  5.30\ \  (\hspace*{2mm} 8.92,-3.62,0.00)&  5.15\ \  (\hspace*{2mm} 7.07,-2.04,0.12) & 7.973\ \ (7.973,0,0)\cr
     3 & 9.204 &  6.00\ \  (\hspace*{2mm} 7.93,-1.94,0.01)&  5.77\ \  (\hspace*{2mm} 7.10,-1.50,0.17)& 8.787\ \ (8.787,0,0)\cr
     4 & ...   &  6.64\ \  (10.47,-3.84,0.01)&  6.53\ \  (\hspace*{2mm} 7.77,-1.94,0.70)& 9.204\ \ (9.204,0,0)\cr
     5 &       &  7.49\ \  (\hspace*{2mm} 8.68,-1.23,0.04)&  6.62\ \  (\hspace*{2mm} 8.66,-2.16,0.12)& ...                 \cr
     6 &       &  7.74\ \  (11.78,-4.06,0.02)&  7.32\ \  (\hspace*{2mm} 8.77,-1.85,0.40)&\cr
     7 &       &  8.37\ \  (10.48,-2.13,0.02)&  7.68\ \  (10.06,-2.52,0.14)&\cr
     8 &       &  9.14\ \  (12.07,-2.95,0.02)&  8.11\ \  (\hspace*{2mm} 8.58,-0.87,0.40)&\cr
     9 &       &  9.26\ \  (11.14,-1.91,0.03)&  8.39\ \  (10.78,-2.96,0.57)&\cr
    10 &       &  ...                        &  8.73\ \  (10.07,-1.62,0.28)&\cr
    11 &       &                             &  9.04\ \  (\hspace*{2mm} 9.46,-0.42,0.00)&\cr
    12 &       &                             &  9.31\ \  (11.10,-1.84,0.05)&\cr
   ... &       &                             &  ...      &\cr
\hline
\end{array}
\nonumber
$
\caption{\small The positive-parity spin-0 energy spectrum for the pure-gluon case (second column),
for the quark-gluon case with 2 left- and 2 right-handed quarks (third column),
with 3 left- and 1 right-handed quark and vice versa (fourth column),
and with 4 left- or 4 right-handed quarks (fifth column).
The 3 numbers in brackets in the third, fourth and fifth column refer to the contributions
from the 3 parts ${\cal H}^G_{\xi=0}$, ${\cal H}^D$, and ${\cal H}^C_{\xi=0}$.
The numerical errors are expected to be smaller than the last digit shown.}
\end{table}
\begin{figure}
\centering
\epsfig{figure=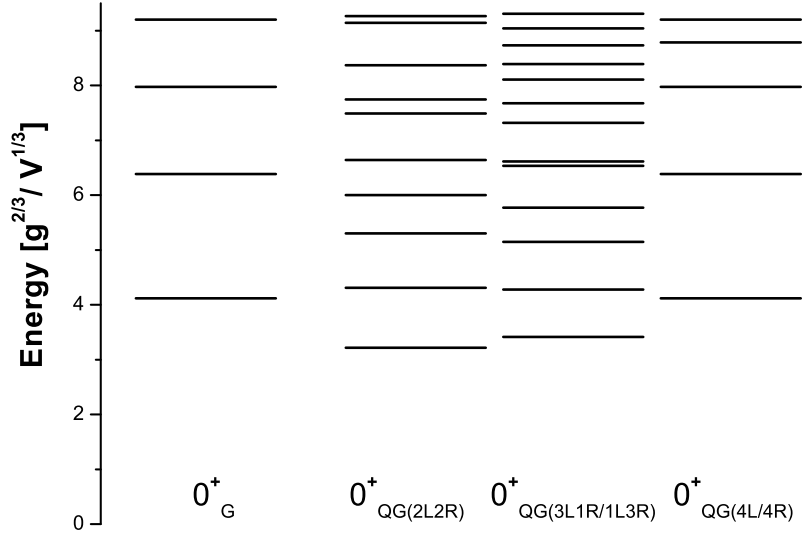,width=120mm}
\caption{\small The energy spectrum of the first positive-parity spin-0 eigenstates
for the pure-gluon case (first spectrum)
and for the quark-gluon case with 2 left-handed and 2 right-handed quarks (second spectrum),
with 3 left-handed and 1 right-handed quark and vice versa (third spectrum),
and with 4 left-handed or 4 right-handed quarks (fourth spectrum).
The numerical errors are expected to be inside the lines.}
\label{energy}
\end{figure}

\section{\large\bf Summary and discussion}
The quantum mechanics of spatially constant $SU(2)$ Yang-Mills fields minimally coupled to Dirac fields
has been investigated as the strong coupling limit of 2-color QCD.
Using the polar decomposition for the spatial components of the original constrained
gauge fields and a corresponding $SU(2)$ phase rotation of the constrained quark fields as in \cite{GKMP},
the corresponding unconstrained Hamiltonian (\ref{unconstrainedH}) and total spin operator (\ref{unconstrainedJ})
has been derived in the present work.
The classical analog (in some equivalent form) of (\ref{unconstrainedH}) has already been obtained in \cite{GKMP}.
The expression for the total spin (\ref{unconstrainedJ}), in particular its quark part $J^Q_i$,
is new in the present work and has important consequences.

Firstly, such as the unconstrained gluon fields can be represented by spin-0 and spin-2 fields,
the unconstrained quark fields are found to carry spin-0 and spin-1, but to continue
to satisfy anti-commutation relations and hence the Pauli-exclusion principle.
The states of $SU(2)$ Dirac-Yang-Mills quantum mechanics can therefore only have integer spin.

Secondly, the expression (\ref{S4}) for $J^Q_i$ in terms of $\widehat{S}^{(4)}$
allows  to determine the correct transformation properties of the unconstrained quark fields
leading to the decomposition (\ref{patransfferm})-(\ref{U_0}).
Together with the well-known principle-axes representation (\ref{patransf}) of the unconstrained gluon field $S$
it leads to a transparent form (\ref{Hphys})-(\ref{Hphysm})
of the unconstrained $SU(2)$-Dirac-Yang-Mills Hamiltonian,
which separates the rotational from the scalar degrees of freedom
and is new to the best of my knowledge. It generalizes the corresponding form of the pure-gluon Hamiltonian,
known already for several decades, and allows to derive the energy spectrum of
the Hamiltonian of $SU(2)$ Dirac-Yang-Mills quantum mechanics.
In order to keep the formulas simple, only the 1-flavor case
has been shown in this work, the generalization of the Hamiltonian (\ref{Hphys})-(\ref{Hphysm})
to two or three flavors is trivial.

As an illustrative example, the energy spectrum has been obtained here for the case of total spin-0
and for 4 massless quarks of one flavor (half-filling), imposing the boundary condition (\ref{fermbc})
of C-symmetry on the quark wave function.
For the case of 2 left- and 2 right-handed quarks, the ground state energy is found to be lowered
in comparison with the pure-gluon case by about $20\%$.
Furthermore, the formation of a quark condensate appears,
in addition to an expected gluon condensate (to be calculated).
Responsible for these features is the part ${\cal H}^D$, which is invariant under the combined
transformation $\phi\rightarrow -\phi$ in the gluon sector and $L \leftrightarrow R$ in the quark sector,
but not invariant under each of them separately.
An energy slightly higher than the ground state, but still considerably lower than that for the pure gluon case,
is obtained, if 3 left- and 1 right-handed quark and vice versa are considered.
Its lowest state could be identified with the sigma-antisigma excitation in our investigation.
The energy spectrum of the third case of 4 left- or 4 right-handed quarks, finally,
is the sum-set of the positive-and negative-parity pure-gluon spectra.
A non-vanishing quark-mass term will cause transitions between these three massless spectra.
Furthermore, it turns out that the boundary condition (\ref{fermbc}) of C-symmetry is the only one
that leads to energies lower than the pure-gluon ground state energy.

Finally, let us remark that, since the fields and states of $SU(2)$ Dirac-Yang-Mills quantum mechanics
all carry integer spin, the expansion of the physical Hamiltonian in the number of spatial derivatives
and coarse graining, developed in \cite{pavel2} for the case of
$SU(2)$ Yang-Mills theory, can be straightforwardly generalized to the case of 2-color-QCD.
One should note that the reduced quarks are not valence quarks, and that baryon number is expected to be related
to topological quantum numbers of field configurations extending over several granulas.
We conjecture here, that for the real case of QCD
with three colors, the situation will be analogous to the 2-color case,
except that baryons now carry half-integer spin and will appear naturally as skyrmions \cite{Skyrme}.

\section*{\large\bf Acknowledgments}

I would like to thank A. Dorokhov and  J. Wambach for their interest and M. Buballa and L. v. Smekal for
discussions on chiral symmetry. Financial support by the LOEWE-Program HIC for FAIR is gratefully acknowledged.

\section*{\large\bf Appendix: Transformation of the quark operators to the intrinsic frame}

The transformation (\ref{patransfferm}) on the fermion fields is chosen such that
$\widehat{S}_i^{(4)}$ and $\widehat{T}_i^{(4)}$ in (\ref{S4a}) and (\ref{T4a}) take the form\footnote{
using the $SO(3)$ identities
$\
\widetilde{R}^T (\chi)\  \widehat{S}_i^{(4)}\ \widetilde{R}(\chi) = R_{ij}(\chi)\ \widehat{S}_j^{(4)}
$
and
$\
\widetilde{R}^T (\chi)\  \widehat{T}_i^{(4)}\ \widetilde{R}(\chi) = R_{ij}(\chi)\ \widehat{T}_j^{(4)}
$
}
\begin{eqnarray}
\widetilde{U}^\dagger (\chi)\  \widehat{S}_i^{(4)}\ \widetilde{U}(\chi)
&=&
R_{ij}(\chi)\ \widetilde{U}^\dagger_0\ \widehat{S}_j^{(4)}\ \widetilde{U}_0
= R_{ij}(\chi)
\left(\begin{array}{c|ccc}
0&0&0&0\cr\hline
0&     &       &    \cr
0&     &  \widehat{S}_j  &    \cr
0&     &       &
\end{array}\right)~,
\nonumber\\
\widetilde{U}^\dagger (\chi)\  \widehat{T}_i^{(4)}\ \widetilde{U}(\chi)
&=&
R_{ij}(\chi)\ \widetilde{U}^\dagger_0\ \widehat{T}_j^{(4)}\ \widetilde{U}_0
= R_{ij}(\chi)
\left(\begin{array}{c|ccc}
0&&\vec{e}_j&\cr\hline
&  0   &    0   &  0  \cr
\vec{e}_j^{\ T}&  0   &  0  &  0  \cr
&  0   &   0    & 0
\end{array}\right)~.\nonumber
\end{eqnarray}
with the $3\times 3$ rotation matrices $(\widehat{S}_i)_{jk} = -i\epsilon_{ijk}$ and
the $3$-dimensional unit vectors $(\vec{e}_i)_j = \delta_{ij}$, and hence
\begin{eqnarray}
J^Q_i =\psi^{\prime \dag}_L \widehat{S}_i^{(4)}\psi^{\prime}_L+
   \psi^{\prime \dag}_R \widehat{S}_i^{(4)}\psi^{\prime}_R=R_{ij}(\chi)\ \widetilde{J}^Q_j~,
\quad\quad\quad\quad
K^Q_i =\psi^{\prime \dag}_L \widehat{T}_i^{(4)}\psi^{\prime}_L+
   \psi^{\prime \dag}_R \widehat{T}_i^{(4)}\psi^{\prime}_R=R_{ij}(\chi)\ \widetilde{K}^Q_j~,
\nonumber
\end{eqnarray}
with the operators (\ref{JKtilde}).
Furthermore, using
\begin{eqnarray}
\widetilde{U}^\dagger (\chi)\left(-S_{ij}(\chi,\phi)\ \tau_i\otimes\tau_j\right)\widetilde{U}(\chi)
&=&
\widetilde{U}^\dagger_0
\left(\begin{array}{cccc}
-\phi_3&0&0&\phi_2-\phi_1\cr
0&\phi_3&-(\phi_1+\phi_2)&0\cr
0&-(\phi_1+\phi_2)&\phi_3&0\cr
\phi_2-\phi_1&0&0&-\phi_3
\end{array}\right)
\widetilde{U}_0\ =
\nonumber\\
&&\!\!\!\!\!\!\!\!\!\!\!\!\!\!\!\!\!\!\!\!\!\!\!\! =
\left(\begin{array}{cccc}
\phi_1+\phi_2+\phi_3&0&0&0\cr
0&\phi_1-(\phi_2+\phi_3)&0&0\cr
0&0&\phi_2-(\phi_3+\phi_1)&0\cr
0&0&0&\phi_3-(\phi_1+\phi_2)
\end{array}\right)\nonumber
\end{eqnarray}
one finds that the minimal coupling part of the Hamiltonian is diagonalised,
\begin{eqnarray}
g \psi^{\prime \dagger}\alpha_i S_{ij} {1\over 2}\tau_j \psi^{\prime} &=&
{g\over 2}\left(\psi^{\prime \dagger}_L,\psi^{\prime \dagger}_R\right)
\left(\begin{array}{cc} -S_{ij} \tau_i\otimes\tau_j &0\cr
0&S_{ij} \tau_i\otimes\tau_j\end{array}\right)
\left(\begin{array}{c} \psi^{\prime}_L\cr \psi^{\prime}_R\end{array}\right)
\nonumber\\
&=&{g\over 2}(\phi_1+\phi_2+\phi_3)
\left(\widetilde{N}^{(0)}_L-\widetilde{N}^{(0)}_R\right)
+{g\over 2}
\sum_{ijk}^{\rm cyclic}\!\! \Bigg[\!\left(\phi_i-(\phi_j+\phi_k)\right)\!
\left(\widetilde{N}^{(i)}_L-\widetilde{N}^{(i)}_R\right)
\Bigg]~.
\nonumber
\end{eqnarray}
with the number-operators (\ref{Ntilde}) of reduced quarks.



\begin{thebibliography}{99}
%
\bibitem{Christ and Lee}
N.H. Christ and T.D. Lee, Phys. Rev. D 22 (1980) 939.
%
\bibitem{GKMP}
S.A. Gogilidze, A.M. Khvedelidze, D. M. Mladenov and H.-P. Pavel,
Phys. Rev. D 57 (1998) 7488.
%
\bibitem{KP1}
A.M. Khvedelidze and H.-P. Pavel,
Phys. Rev. D 59 (1999) 105017.
%
\bibitem{KMPR}
A.M. Khvedelidze, D. M. Mladenov, H.-P. Pavel, and G. R\"opke,
Phys. Rev. D 67 (2003) 105013.
%
\bibitem{pavel2}
H.-P. Pavel, Phys. Lett. B 685 (2010) 353.
%
\bibitem{Luescher and Muenster}
M. L\"uscher and G. M\"unster, Nucl. Phys. B232 (1984) 445.
%
\bibitem{Savvidy}
G. K. Savvidy, Phys. Lett. 159B (1985) 325.
%
\bibitem{Simon}
Yu. Simonov, Sov. J. Nucl. Phys. 41 (1985) 835.
%
\bibitem{Koller and van Baal}
J. Koller and P. van Baal, Nucl. Phys. B273 (1986) 387;
Nucl. Phys. B302 (1988) 1;
P. van Baal and J. Koller,
Ann. of Phys. (N.Y.) 174 (1987) 299.
%
\bibitem{Weisz}
P. Weisz and V. Ziemann, Nucl. Phys. B284 (1987) 157.
%
\bibitem{KP2}
A. Khvedelidze and H.-P. Pavel,
Phys. Lett. A 267 (2000) 96;
%
A. Khvedelidze, H.-P. Pavel, and  G. R\"opke,
Phys. Rev. D 61 (2000) 025017.
%
\bibitem{pavel1}
H.-P. Pavel, Phys. Lett. B 648 (2007) 97.
%
\bibitem{van Baal}
P. van Baal, Nucl. Phys. B307 (1988) 274.
%
\bibitem{Michael}
J. Kripfganz and C. Michael, Phys. Lett. 209B (1988) 77;
Nucl. Phys. B314 (1989) 25.
%
\bibitem{Skyrme}
T.H.R. Skyrme, Proc. Royal Soc. A260 (1961) 127,
Nucl. Phys. 31 (1962) 556.
%
\end{thebibliography}
\end{document}